\documentclass[showpacs,aps,pre,superscriptaddress]{revtex4}
\usepackage{amsmath}
\usepackage{graphicx}
\usepackage{amsfonts}

\begin{document}

\title{Symmetry breaking, coupling management, and localized modes in
dual-core discrete nonlinear-Schr\"{o}dinger lattices}
\author{H. Susanto}
\affiliation{School of Mathematical Sciences, University of Nottingham, University Park,
Nottingham, NG7 2RD, United Kingdom}
\author{P. G.\ Kevrekidis}
\affiliation{Department of Mathematics and Statistics, University of Massachusetts,
Amherst MA 01003-4515, USA}
\author{F.\ Kh.\ Abdullaev}
\affiliation{Physical-Technical Institute of the Academy of Sciences, 700084,
Tashkent-84, G.Mavlyanov str.,2-b, Uzbekistan}
\author{Boris A.\ Malomed}
\affiliation{Department of Physical Electronics, School of Electrical Engineering,
Faculty of Engineering, Tel Aviv University, Tel Aviv 69978, Israel}

\begin{abstract}
We introduce a system of two linearly coupled discrete nonlinear Schr\"{o}%
dinger equations (DNLSEs), with the coupling constant subject to a rapid
temporal modulation. The model can be realized in bimodal Bose-Einstein
condensates (BEC). Using an averaging procedure based on the multiscale
method, we derive a system of averaged (autonomous) equations, which take
the form of coupled DNLSEs with additional nonlinear coupling terms of the
four-wave-mixing type. We identify stability regions for fundamental onsite
discrete symmetric solitons (single-site modes with equal norms in both
components), as well as for two-site in-phase and twisted modes, the
in-phase ones being completely unstable. The symmetry-breaking bifurcation,
which destabilizes the fundamental symmetric solitons and gives rise to
their asymmetric counterparts, is investigated too. It is demonstrated that
the averaged equations provide a good approximation in all the cases. In
particular, the symmetry-breaking bifurcation, which is of the pitchfork
type in the framework of the averaged equations, corresponds to a Hopf
bifurcation in terms of the original system.
\end{abstract}

\maketitle

\section{Introduction}

The problem of the influence of rapidly varying perturbations on solitons
belongs to the general topic of the ``soliton management" \cite{book}. Along
with other problems of this kind, this one has drawn considerable attention.
The motivation is to investigate new nontrivial dynamics that may be induced
by rapidly varying perturbations, and predict, in this way, new types of
solitons in such systems. In terms of mechanical systems, with few degrees
of freedom, the closest counterparts of this setting are presented by the
Kapitza pendulum \cite{Landau} and the motion of a charged particle in
rapidly oscillating electromagnetic fields. In either case, the averaged
dynamics is governed by an effective potential, different from the original
one, whose fixed points may give rise to novel dynamical states. For the
discrete nonlinear Schr\"{o}dinger (DNLS) equation, problems of this kind
(``rapid management") were considered for a variable second-order discrete
dispersion \cite{AM}, and for the variable nonlinearity \cite{ATMK}. The
respective averaged equations take the form of a nonlocal generalization of
the DNLSE, which gives rise to new types of localized breathers. Recently,
the existence of a soliton in the DNLSE with an external drive has been
demonstrated in Ref. \cite{GAS}, and the influence of a temporal delay in
the onsite nonlinearity on the self-trapping of discrete solitons was
studied in Ref. \cite{new}.

The above-mentioned examples pertain to single-component discrete systems. A
natural physically relevant generalization is to explore the influence of
rapidly varying strong perturbations on the dynamics of discrete vectorial
(two-component) solitons. In this work,we introduce a system of two linearly
coupled DNLSEs with a rapidly varying coupling parameter. This example,
which is interesting in its own right, is a straightforward model of
Bose-Einstein condensates (BECs) confined in two parallel tunnel-coupled
cigar-shaped traps, combined with a deep optical lattice (OL), in a case
when the linear-coupling parameter is subject to the temporal modulation
\cite{AK,SK}. Another implementation of the model is possible in terms of
the binary BEC (also loaded into a deep OL) with two components linearly
coupled by a resonant electromagnetic wave \cite{Saito,Susanto}, whose
amplitude may also be periodically modulated in time \cite{Susanto}. In the
latter case, the coupled system of the Gross-Pitaevskii equations also
contains nonlinear-interaction terms, accounting for collisions between
atoms belonging to the different species. However, the coefficient in front
of the latter terms may be effectively switched off by means of the Feshbach
resonance affecting the inter-species collisions (see, e.g., Ref. \cite%
{Italy}), which we assume below. Our goal is to derive effective averaged
equations approximating this model, and thus predict discrete solitons in
it--both symmetric ones, and asymmetric states generated by the \textit{%
symmetry-breaking bifurcation} in the dual-core system. The analysis of this
bifurcation in ``unmanaged" dual-core DNLSE systems, with a constant
linear-coupling constant, was developed in Refs. \cite{TM} and \cite{herr07}.

The paper is structured as follows. The model is introduced in
Section II, where we also derive the approximation based on the
averaged equations. Symmetric discrete solitons, of both single-site
and two-site types (in-phase and twisted ones, in the latter case),
are investigated in Section III. In Section IV, we explore the
symmetry-breaking bifurcation, which destabilizes the single-site
symmetric solitons, giving rise to their asymmetric counterparts.
Conclusions are formulated in Section V.

\section{The model and averaged equations}

Following the above discussion, we introduce the following system of
linearly coupled DNLSEs, written in a scaled form,
\begin{subequations}
\label{sys1}
\begin{eqnarray}
i\frac{d}{dt}u_{n}+\kappa _{1}(u_{n+1}-u_{n-1})+\gamma
_{0}|u_{n}|^{2}u_{n}+f(t)v_{n} &=&0, \\
i\frac{d}{dt}v_{n}+\kappa _{2}(v_{n+1}+v_{n-1})+\gamma
_{0}|v_{n}|^{2}v_{n}+f(t)u_{n} &=&0,
\end{eqnarray}
for amplitudes $u_{n}(t)$ and $v_{n}(t)$ of the two tunnel-coupled BECs
trapped in the deep OL \cite{TM}, with time-dependent coupling coefficient $%
f(t)$. Here, we consider the case of the strong high-frequency temporal
modulation, $f(t)=f_{0}+f_{1}\sin (\Omega t),$with $f_{1}\sim \Omega \gg 1$.

Different methods can be used to derive averaged equations for slowly
varying fields, such as the multi-scale expansion \cite{Kath} and the
Kapitza method \cite{Landau}. In the following, we derive the averaged
system by way of the former technique. To this end, we first define $C(\tau
)=\widetilde{f_{1}}\sin (\tau ),$ where $\tau \equiv \Omega t$ and $%
\widetilde{f_{1}}\equiv f_{1}/\Omega $, so that $\widetilde{f_{1}}=\mathcal{O%
}(1)$. To remove the secular part of the coupling terms, we use a linear
transformation,
\end{subequations}
\begin{equation}
\left(
\begin{array}{c}
u_{n} \\
v_{n}%
\end{array}%
\right) =\left(
\begin{array}{cc}
\cos (C_{-1}(\tau )) & -i\sin (C_{-1}(\tau )) \\
-i\sin (C_{-1}(\tau )) & \cos (C_{-1}(\tau ))%
\end{array}%
\right) \left(
\begin{array}{c}
\phi _{n} \\
\psi _{n}%
\end{array}%
\right) ,  \label{trans}
\end{equation}%
where, in the case of the general time modulation, $C_{-1}(\tau
)=\int_{0}^{\tau }C(\tau ^{\prime })\,d\tau ^{\prime
}-P^{-1}\int_{0}^{P}\int_{0}^{\tau }C(\tau ^{\prime })\,d\tau ^{\prime
}\,d\tau ,$ and $P$ is the period of $C(\tau )$. For the periodic coupling
defined above, $C_{-1}=-\widetilde{f_{1}}\cos (\Omega t)$, and $P=2\pi $.

Substituting $\phi _{n}=\Phi _{n}(t)+\mathcal{O}(1/\Omega )$ and $\psi
_{n}=\Psi _{n}(t)+\mathcal{O}(1/\Omega )$ and averaging over the fast time
variable, $\tau $, it is straightforward to derive the following averaged
equation for $\Phi _{n}$ and $\Psi _{n}$, 
\begin{subequations}
\label{sys2}
\begin{eqnarray}
i\frac{d}{dt}\Phi _{n}+\kappa _{1}(\Phi _{n+1}+\Phi _{n-1})+\frac{\gamma _{0}%
}{4}\left[ (3+\sigma )|\Phi _{n}|^{2}+2(1-\sigma )|\Psi _{n}|^{2}\right]
\Phi _{n}+\frac{\gamma _{0}}{4}(\sigma -1)\Phi _{n}^{\ast }\Psi
_{n}^{2}+f_{0}\Psi _{n} &=&0, \\
i\frac{d}{dt}\Psi _{n}+\kappa _{2}(\Psi _{n+1}+\Psi _{n-1})+\frac{\gamma _{0}%
}{4}\left[ 2(1-\sigma )|\Phi _{n}|^{2}+(3+\sigma )|\Psi _{n}|^{2}\right]
\Psi _{n}+\frac{\gamma _{0}}{4}(\sigma -1)\Psi _{n}^{\ast }\Phi
_{n}^{2}+f_{0}\Phi _{n} &=&0,
\end{eqnarray}%
%
%
%
where $\sigma =P^{-1}\int_{0}^{P}\cos (4C_{-1}(\tau ))\,d\tau .$ For the
form of the periodic modulation adopted above,
\end{subequations}
\begin{equation}
\sigma =\left( 2\pi \right) ^{-1}\int_{0}^{2\pi }\cos \left( 4\widetilde{%
f_{1}}\cos (\tau )\right) d\tau \equiv J_{0}\left( 4\widetilde{f_{1}}\right)
,  \label{Friedrich}
\end{equation}%
where $J_{0}$ is the Bessel function. Note that the nonlinear coupling terms
generated by the averaging method in Eqs. (\ref{sys2}) are sensitive to the
phase difference between the two fields, i.e., these terms are of the
four-wave-mixing type.

\section{Symmetric localized modes}

\subsection{The general approach}

To check the accuracy of our averaging analysis, in the following we explore
the existence and stability of discrete solitons predicted by Eqs. (\ref%
{sys1}) and (\ref{sys2}). First, we consider stationary solutions which are
symmetric with respect to the coupled subsystems, i.e., with $\Phi _{n}=\Psi
_{n}$.

Using the Newton-Raphson continuation method, we look for stationary
solutions to Eqs.\ (\ref{sys2}). In this case, the solution is sought for in
the form of $\Phi _{n}(t)=e^{\Lambda t}U_{n}$ and $\Psi _{n}(t)=e^{\Lambda
t}V_{n}$, where the frequency is assumed to be fixed as $\Lambda =1$, by
means of an obvious rescaling. Substituting this in Eqs. (\ref{sys2}), we
arrive at the following equations for the real-valued functions, ${U}_{n}$
and ${V}_{n}$: 
\begin{subequations}
\label{sys3}
\begin{eqnarray}
-\Lambda U_{n}+\kappa _{1}(U_{n+1}+U_{n-1})+\frac{1}{4}\gamma _{0}\left(
(3+\sigma )U_{n}^{2}+(1-\sigma )V_{n}^{2}\right) U_{n}+f_{0}V_{n} &=&0, \\
-\Lambda V_{n}+\kappa _{2}(V_{n+1}+V_{n-1})+\frac{1}{4}\gamma _{0}\left(
(1-\sigma )U_{n}^{2}+(3+\sigma )V_{n}^{2}\right) V_{n}+f_{0}U_{n} &=&0.
\end{eqnarray}%
%

To test the stability of the stationary solutions, we substitute the usual
expression for a weakly perturbed solution, $\Phi _{n}(t)=e^{\Lambda
t}\left( U_{n}+e^{i\lambda t}\tilde{u}_{n}\right) $ and $\Psi
_{n}(t)=e^{\Lambda t}\left( V_{n}+e^{i\lambda t}\tilde{v}_{n}\right) $, into
Eqs.\ (\ref{sys2}) and perform the linearization of the equations, arriving
at the eigenvalue problem for perturbation frequencies $\lambda $,%
\end{subequations}
\begin{subequations}
\label{sta}
\begin{eqnarray}
\kappa _{1}(\tilde{u}_{n+1}+\tilde{u}_{n-1})+\frac{\gamma _{0}}{4}\left[
(3+\sigma )U_{n}^{2}(2\tilde{u}_{n}+\tilde{u}_{n}^{\ast })+(1-\sigma )\left(
V_{n}^{2}(2\tilde{u}_{n}-\tilde{u}_{n}^{\ast })+2U_{n}V_{n}\tilde{v}%
_{n}^{\ast }\right) \right] +f_{0}\tilde{v}_{n} &=&\left( \lambda +\Lambda
\right) \tilde{u}_{n}, \\
\kappa _{2}(\tilde{v}_{n+1}+\tilde{v}_{n-1})+\frac{\gamma _{0}}{4}\left[
(3+\sigma )V_{n}^{2}(2\tilde{v}_{n}+\tilde{v}_{n}^{\ast })+(1-\sigma )\left(
U_{n}^{2}(2\tilde{v}_{n}-\tilde{v}_{n}^{\ast })+2U_{n}V_{n}\tilde{u}%
_{n}^{\ast }\right) \right] +f_{0}\tilde{u}_{n} &=&\left( \lambda +\Lambda
\right) \tilde{v}_{n}.
\end{eqnarray}%
%
The discrete soliton is stable if all eigenvalues $\lambda $ have\ $\mathrm{%
Im}(\lambda )=0$.

\subsection{Fundamental discrete solitons (single-site modes)}

We start with the consideration of the existence and stability of a
symmetric single-site mode, which, in the anticontinuum limit of
$\kappa _{1}=\kappa _{2}=0$, is given by $U_{n}=V_{n}=\delta
_{n,n_{0}}$. In the top left panel of Fig. \ref{fig2} we present
numerical results for the stability of this mode in the ($\sigma
,f_{0}$) plane, for a particular value of the lattice coupling
constants, $\kappa _{1}=\kappa _{2}=0.2$. Represented in colors is
the largest imaginary part of the eigenvalues, the stability region
being left black.

\begin{figure}[tbph]
\includegraphics[width=6cm,angle=0,clip]{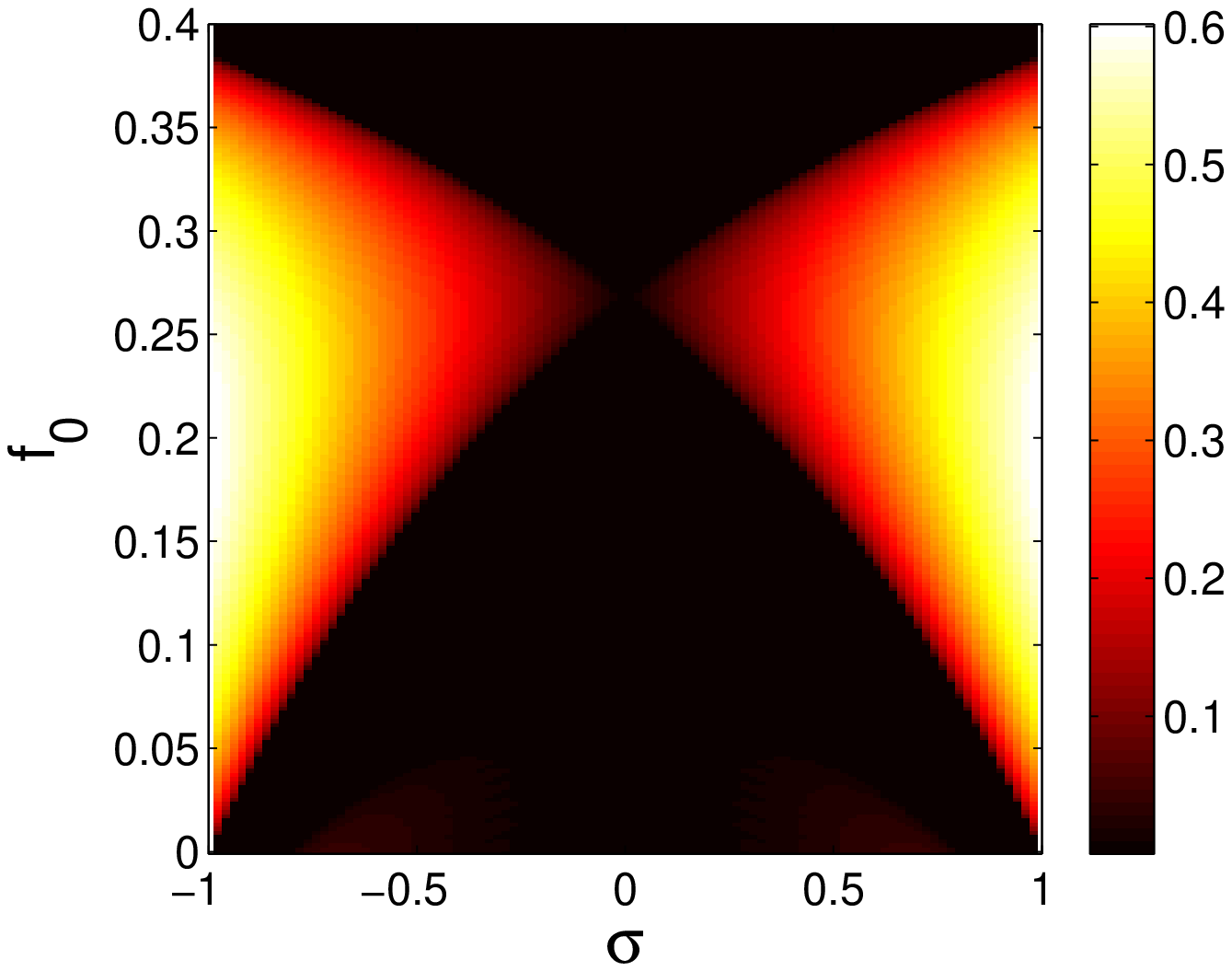} %
\includegraphics[width=6cm,angle=0,clip]{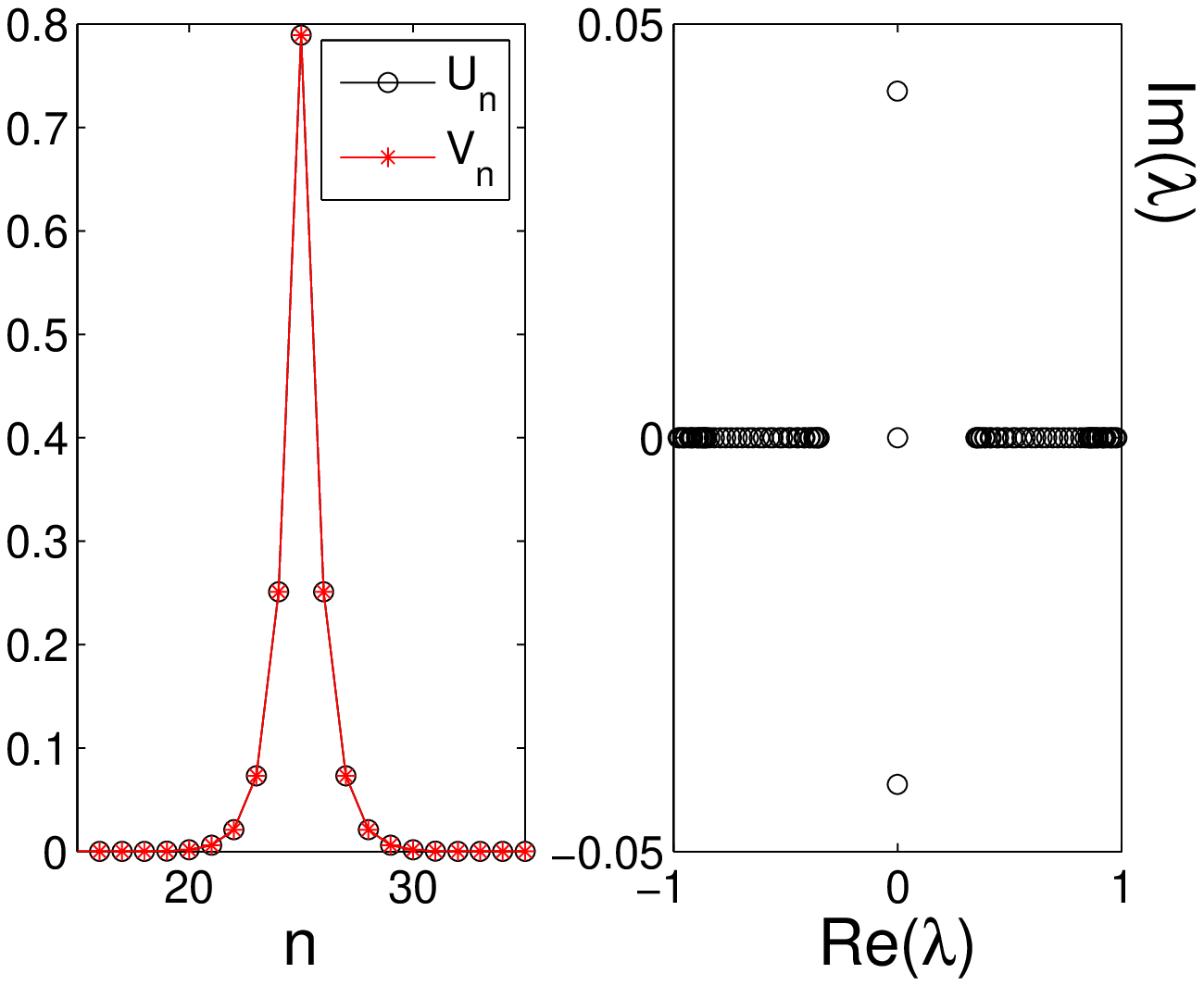} %
\includegraphics[width=6cm,angle=0,clip]{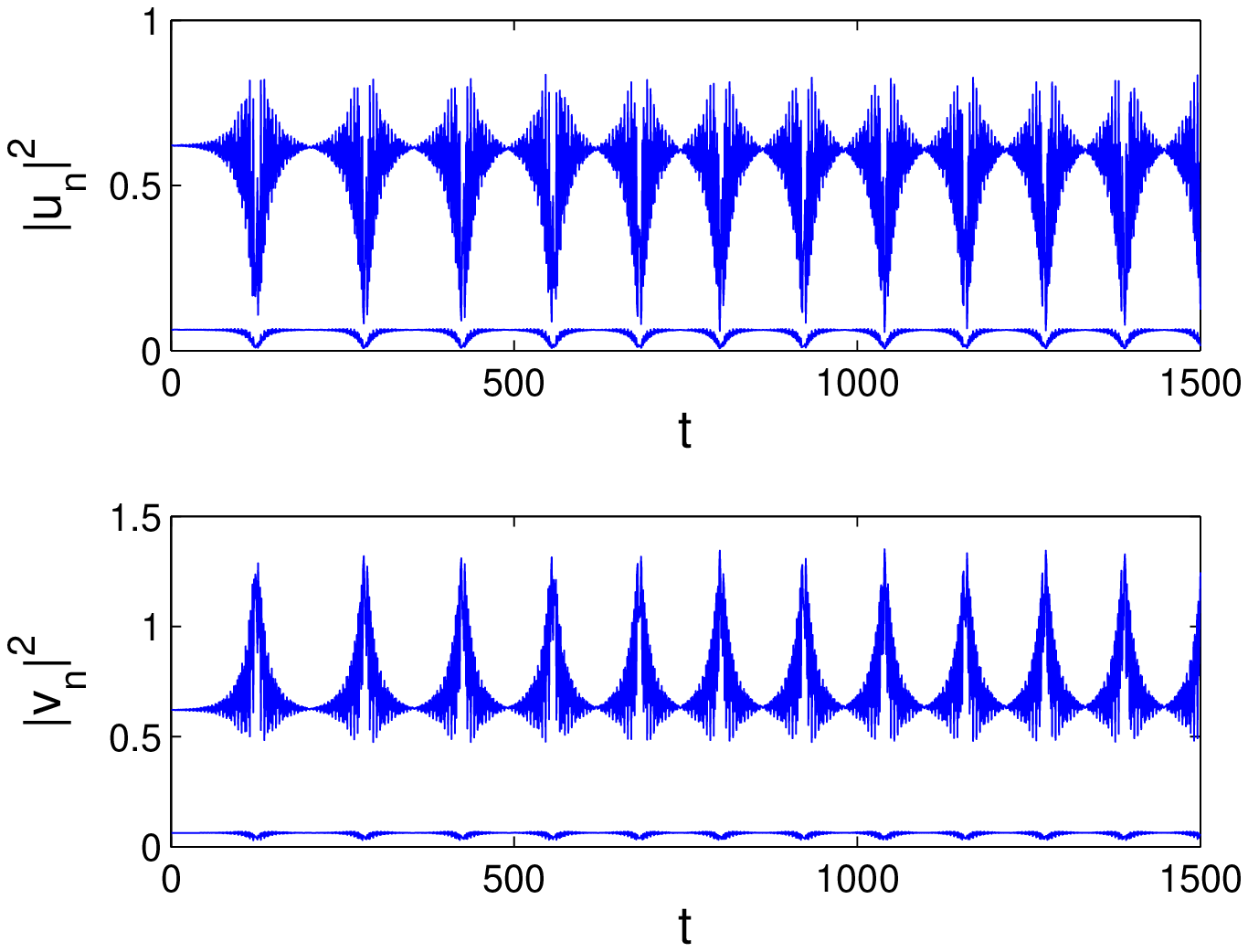} %
\includegraphics[width=6cm,angle=0,clip]{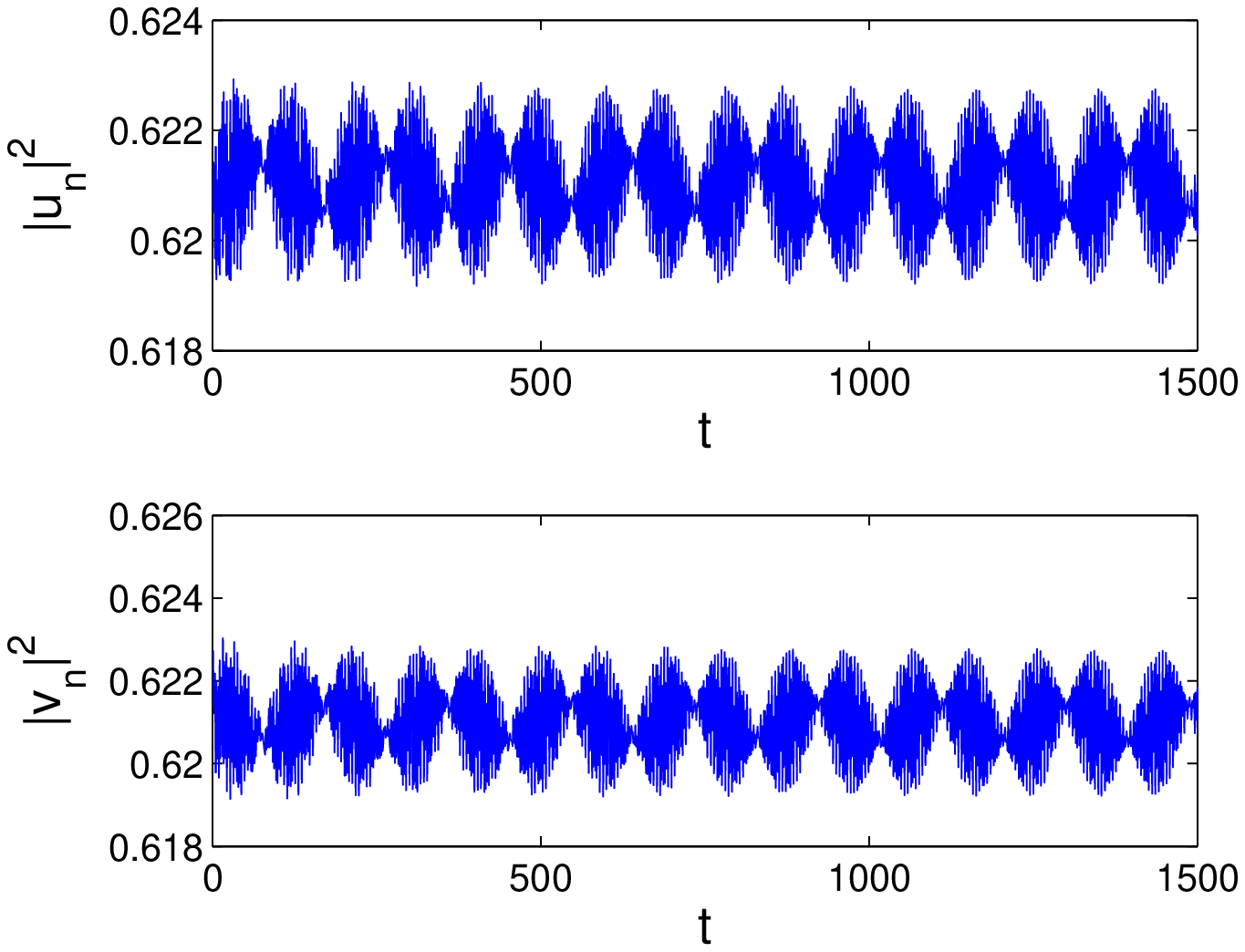}
\caption{(Color online) The top top left panel shows the
(in)stability regions for the onsite discrete solitons in the
$(\protect\sigma ,f_{0})$ plane with $\protect\kappa
_{1}=\protect\kappa _{2}=0.2$. The top right panel depicts a generic
example of the soliton's profile and its stability spectrum at
$f_{1}=5.4$. The bottom panels shows the evolution of the two
components $|u_{n}|^{2}$ and $|v_{n}|^{2}$, at $n=25,26$ (left) and
$n=25$ (right), for $f_{1}=5.4\,$and $5.6$, as found from
simulations of the underlying equations (\protect\ref{sys1}), which
indicates a perfect agreement with the threshold value predicted by
the averaged equation.} \label{fig2}
\end{figure}

As seen in the top left panel, for $f_{0}=0.25$, the stability window of the
localized mode is $-0.1072<\sigma <0.1084$. Using Eq. (\ref{Friedrich}), we
conclude that the values of $f_{1}$ supporting the stable onsite mode are
given by $5.5<f_{1}<6.54$ for $\Omega =10$. In the top right panel we
present the profile of the discrete soliton at $f_{1}=5.4$ and its stability
spectrum. The solution is unstable due to a collision of a pair of
eigenvalues at the origin, resulting in a doublet of purely imaginary
eigenvalues.

To check the accuracy of the predictions provided by the averaged equations (%
\ref{sys3}), we use the static solution obtained from these equations as
initial conditions for simulations of underlying equations (\ref{sys1}),
inverting transformation (\ref{trans}) for this purpose. Shown in the left
bottom panel of Fig.\ \ref{fig2} is the generated evolution of the discrete
soliton, just near the stability border shown in the top right panel of the
same figure. One can see that the unstable soliton tends to rearrange itself
into a different time-periodic localized solution. We will see later that in
the averaged equation (\ref{sys2}) this corresponds to an asymmetric mode.
On the other hand, in the right bottom panel we show the evolution of a
stable onsite soliton at $f_{1}=5.6$, from which one can conclude that our
averaged equations accurately predict the stability threshold, in comparison
with Eqs. (\ref{sys1}).

\subsection{Two-site modes}

Proceeding from the fundamental onsite solitons (single-site modes) to the
consideration of multi-site ones, a natural object is an in-phase two-site
state, i.e., an\ intersite discrete soliton. However, our numerical analysis
(not shown here) reveals that this mode is unstable, in terms of the
averaged equations, for all parameter values. Explicit simulations of the
evolution of the counterpart of this state within the framework of the
original equations (\ref{sys1}) confirm its instability.\

Next, we consider an out-of-phase fundamental two-site state, alias a\
\textit{twisted} localized state, which, in the anticontinuum limit, is
seeded by ansatz $U_{n}=V_{n}=\delta _{n,n_{0}}-\delta _{n,n_{0}+1}$.
Results of the numerical analysis of this family of odd discrete solitons
are presented in Fig. \ref{fig2a}. With $\kappa _{1}=\kappa _{2}=0.05$ and $%
f_{1}=4$, which corresponds to $\sigma =0.455$, the mode is unstable in the
range of $0.223<f_{0}<0.469$. In the left and right bottom panels of the
figure, we show the evolution of the twisted mode for $f_{0}=0.22$ and $%
f_{0}=0.23$, respectively. Again we observe a good agreement between the
stability threshold predicted by averaged equations (\ref{sys2}) and the
original system, Eqs. (\ref{sys1}).

\begin{figure}[tbph]
\includegraphics[width=6cm,angle=0,clip]{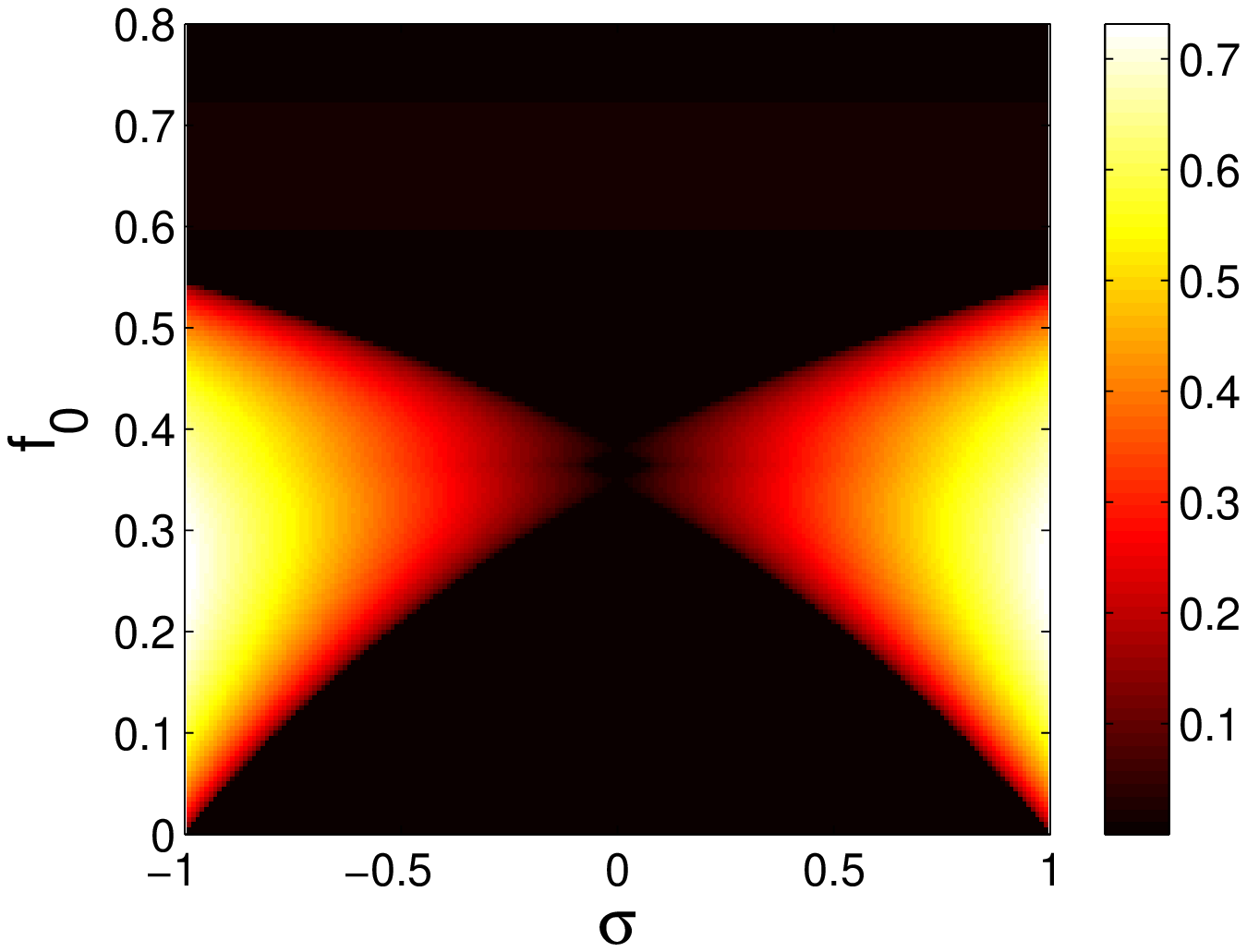} %
\includegraphics[width=6cm,angle=0,clip]{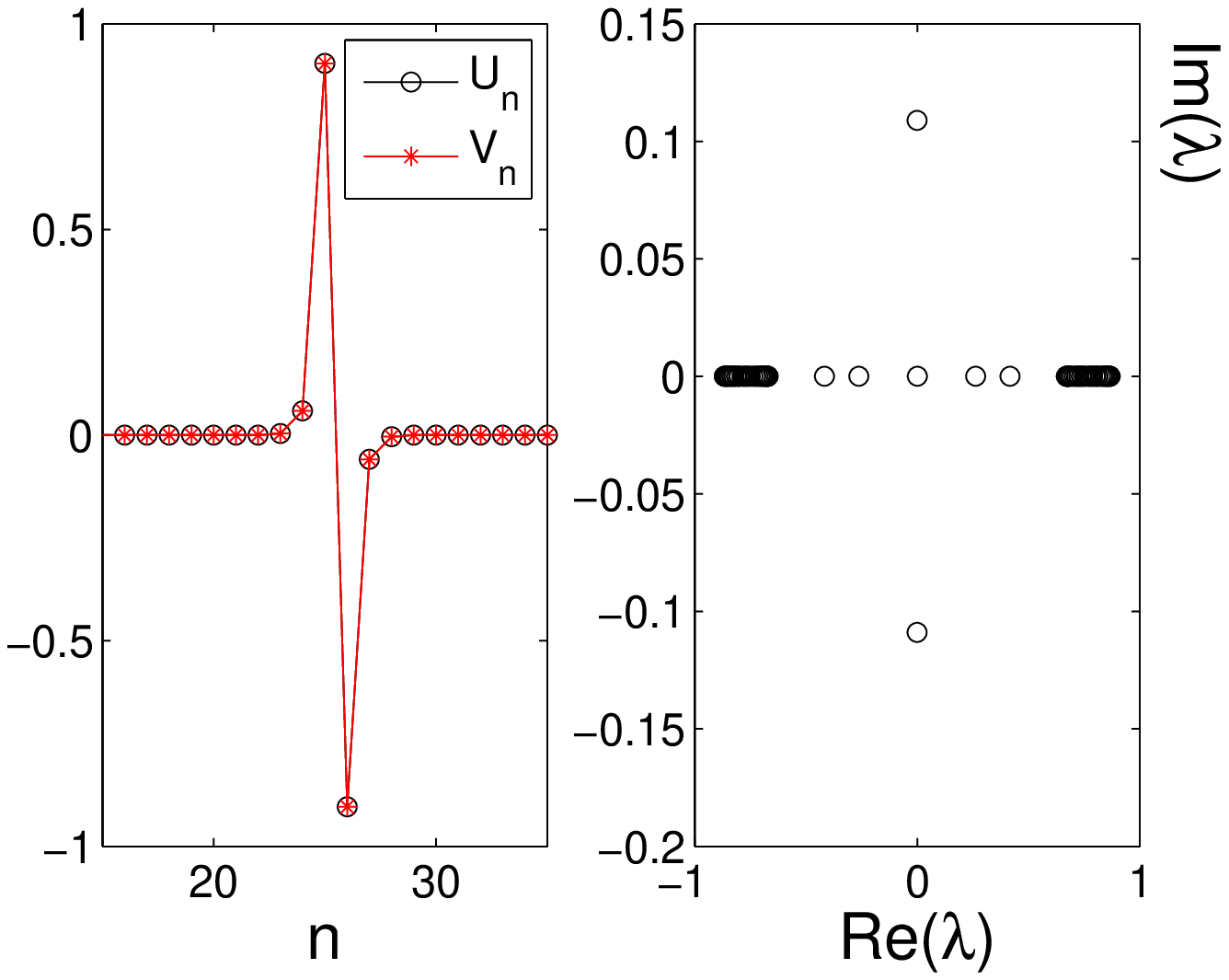} %
\includegraphics[width=6cm,angle=0,clip]{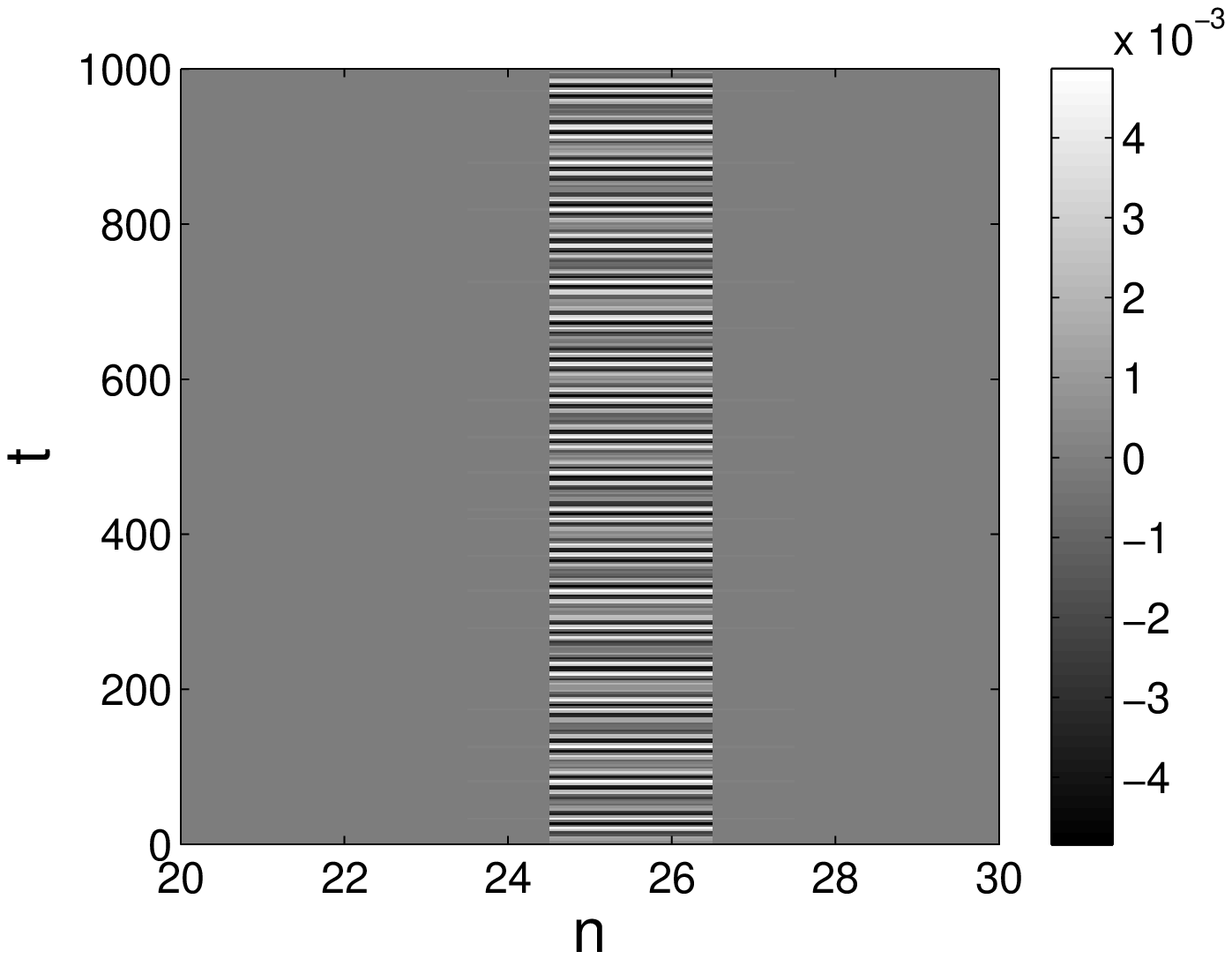} %
\includegraphics[width=6cm,angle=0,clip]{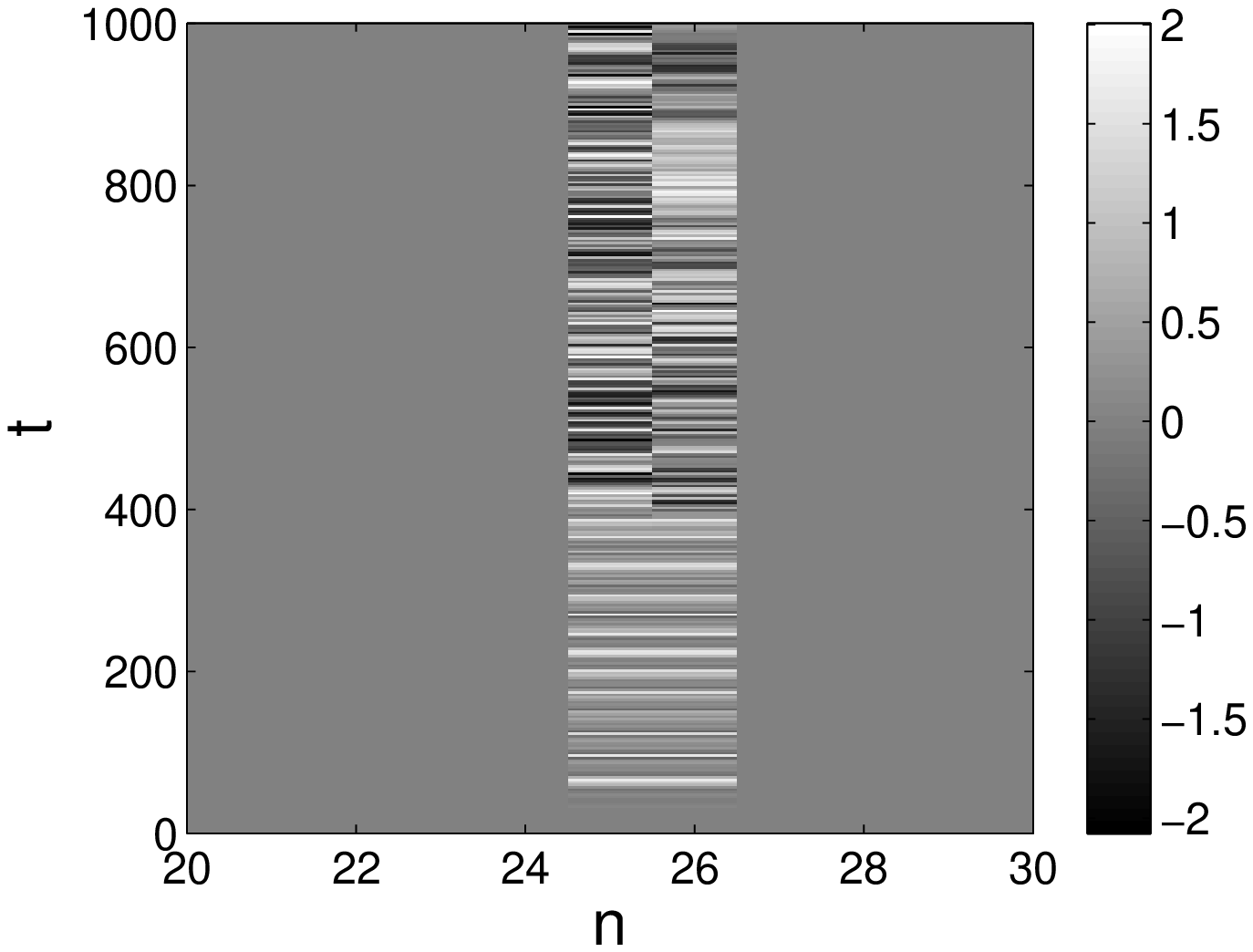}
\caption{(Color online) The top left panel is the same as in Fig.\ \protect
\ref{fig2}, but for a twisted mode with $\protect\kappa _{1}=\protect\kappa %
_{2}=0.05$. The top right panel shows the profile of the mode and
its stability spectrum for $f_{1}=4$ and $f_{0}=0.23$. Bottom panels
show the evolution of stable and an unstable modes at $f_{0}=0.22$
and $\,0.23$, respectively.}
\label{fig2a}
\end{figure}

\section{Symmetry breaking: Pitchfork and Hopf bifurcations}

As reported in Refs.\ \cite{TM} and \cite{herr07}, linearly coupled
two-component systems, which are similar to the present one, give rise to a
symmetry-breaking bifurcation, which generates asymmetric solitons, while
destabilizing symmetric fundamental ones. The analysis of the present model
reveals that a similar bifurcation occurs in averaged equations (\ref{sys2}%
), and it generates the respective asymmetric mode.
Presented in the top left panel of Fig.\ \ref{fig3} is the (in)stability
region of the asymmetric mode for $\kappa _{1}=\kappa _{2}=0.2$. In this
panel, the white curve designates the \textit{bifurcation line}, below which
the asymmetric mode exists. The analysis demonstrates that the bifurcation
in the averaged system (\ref{sys2}) is of the \textit{pitchfork} type, like
in the recently studied related models \cite{TM,herr07}).
As a particular example, we depict in the top right panel of Fig.\ \ref{fig3}
the solution and the stability spectral plane for an unstable asymmetric
onsite mode with $f_{0}=0.23$ and $f_{1}=7$, which corresponds to $\sigma
=-0.19$.

\begin{figure}[tbph]
\includegraphics[width=6cm,angle=0,clip]{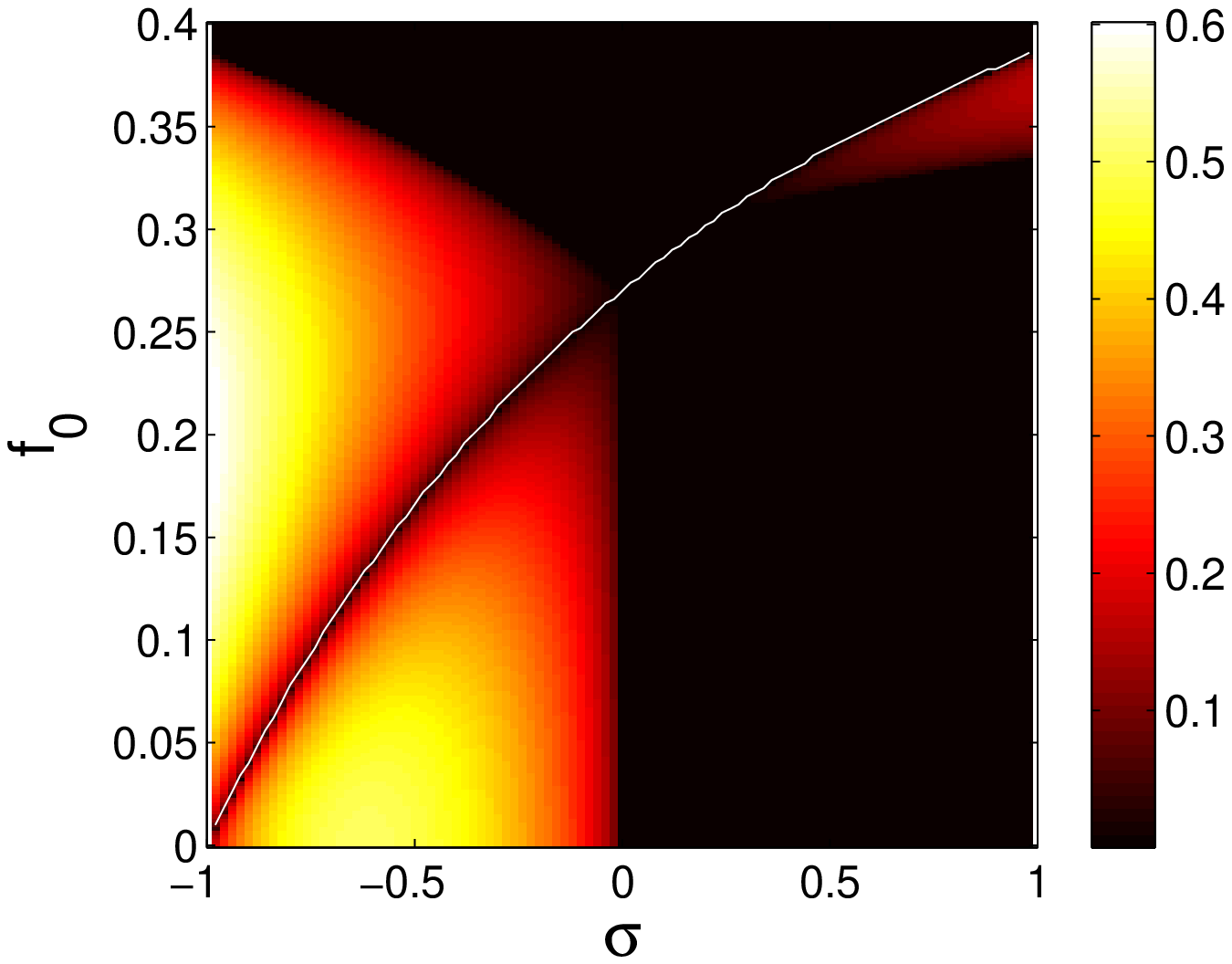} %
\includegraphics[width=6cm,angle=0,clip]{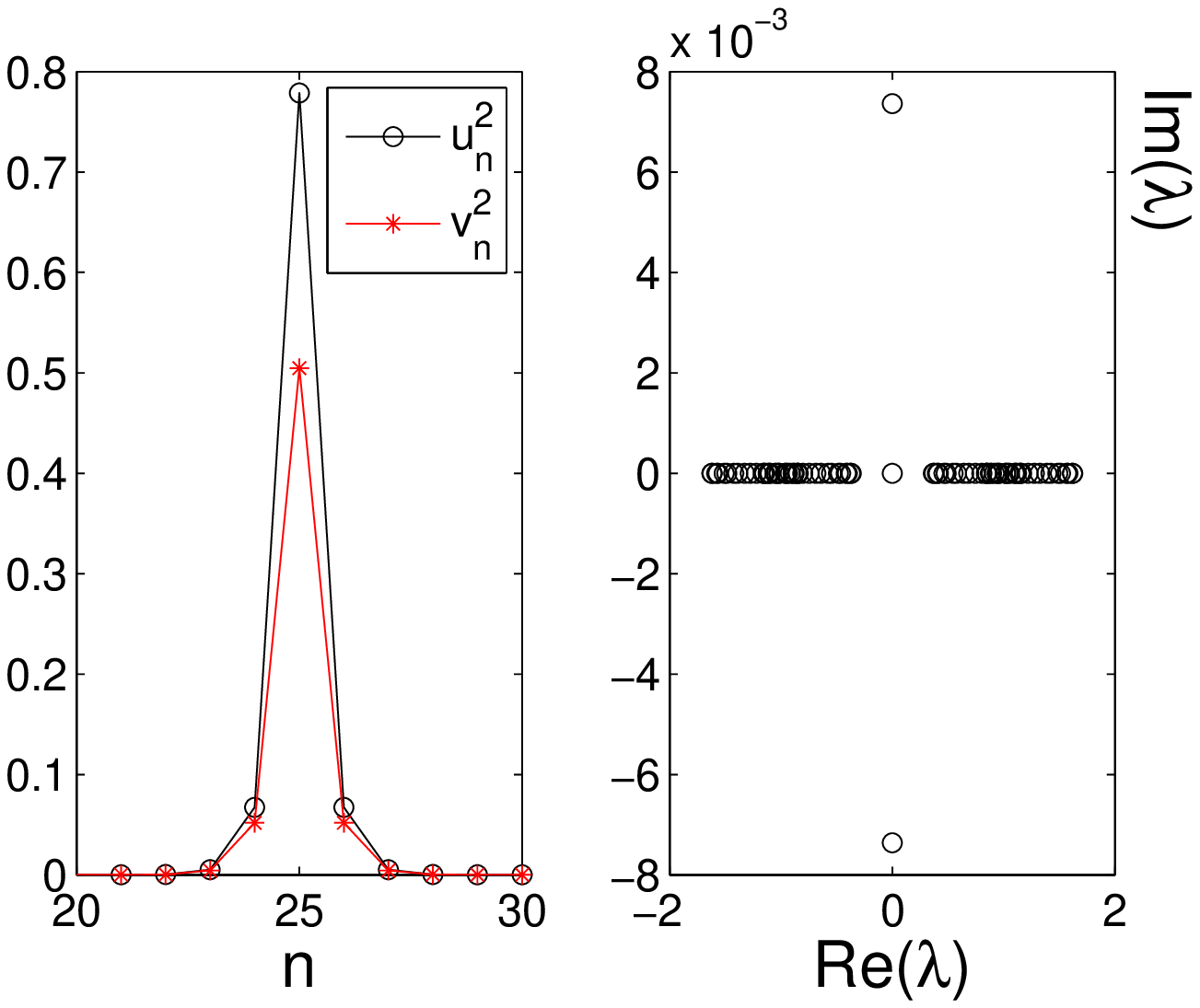} %
\includegraphics[width=6cm,angle=0,clip]{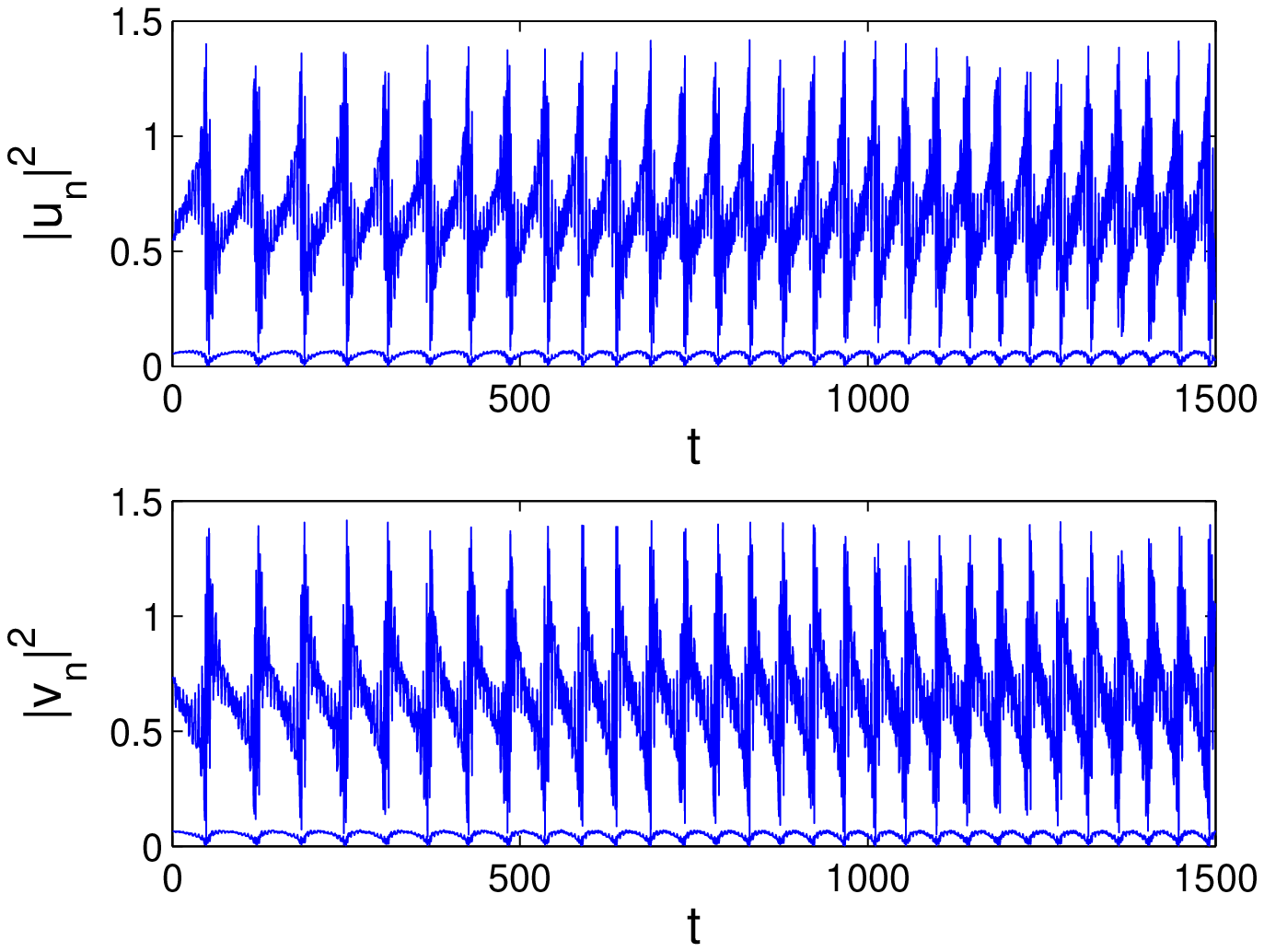} %
\includegraphics[width=6cm,angle=0,clip]{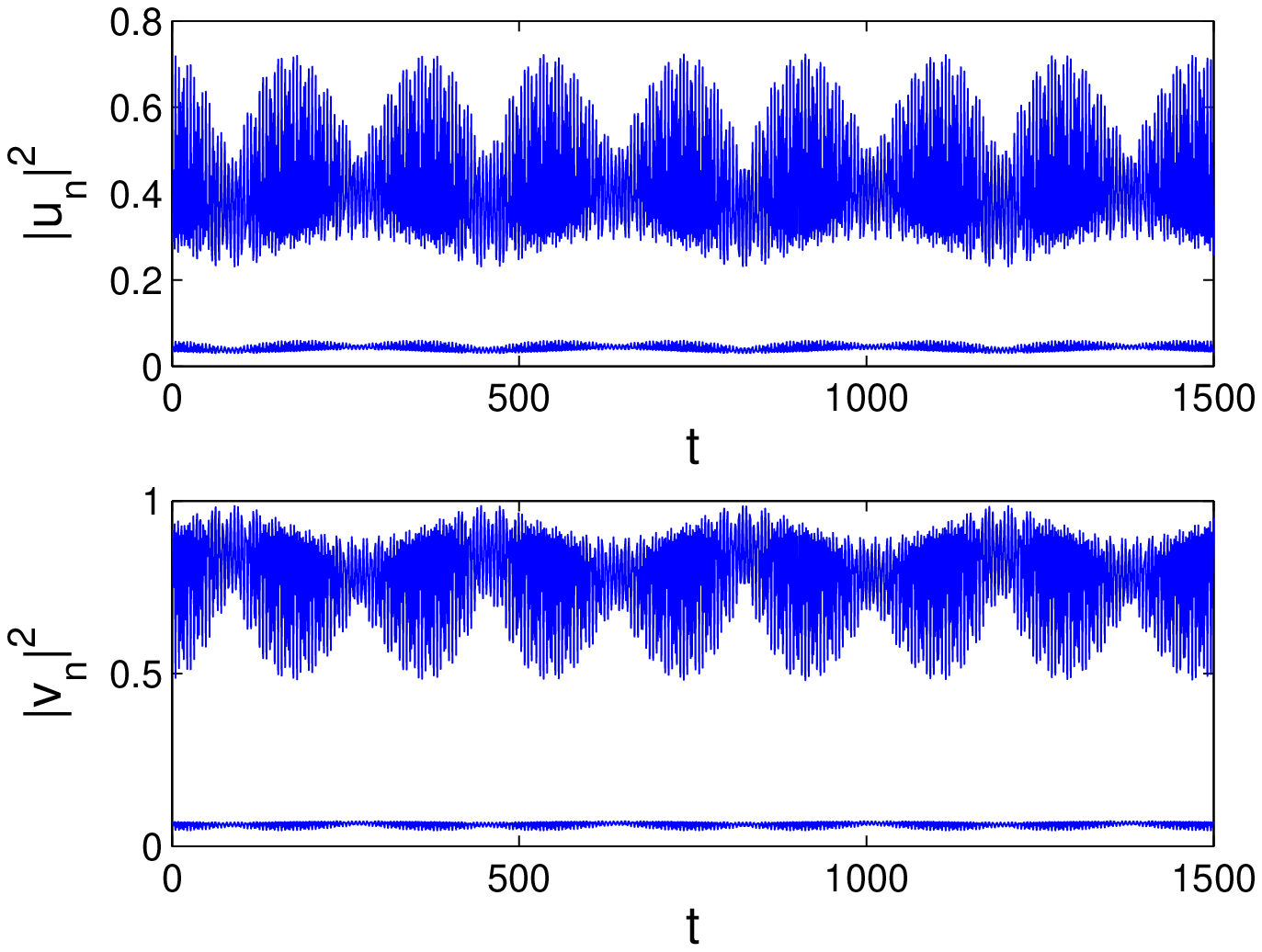}
\caption{(Color online) The same as in Fig.\ \protect\ref{fig2}, but
for the asymmetric mode, generated by the symmetry-breaking
bifurcation, which happens along the while curve in the top left
panel. See the text for further details.}
\label{fig3}
\end{figure}

It is worthy to note that the stability border of the asymmetric mode in the
top left panel exactly corresponds to $\sigma =0$, the asymmetric discrete
soliton being stable at $\sigma >0$. This observation may be explained by
the fact that the self-phase-modulation nonlinear terms in Eqs. (\ref{sys2})
dominate over their cross-phase-modulation counterparts just at $\sigma >0$.

Before we proceed to checking the accuracy of the averaged equations, it is
necessary to make the following notice concerning transformation (\ref{trans}%
): if $|\Phi _{n}(t)|^{2}$ and $|\Psi _{n}(t)|^{2}$ are static, then $%
|u_{n}(t)|^{2}$ and $|v_{n}(t)|^{2}$ will also be static, provided that $%
U_{n}$ and $V_{n}$ are symmetric. On the other hand, if $U_{n}$ and $V_{n}$
are asymmetric, the transformation yields $|u_{n}(t)|^{2}$ and $%
|v_{n}(t)|^{2}$ oscillating periodically in time. Therefore, the \textit{%
pitchfork bifurcation} in the averaged equations (\ref{sys3}) corresponds to
a \textit{Hopf bifurcation} in the original system (\ref{sys1}).

Taking the above-mentioned fact into regard, an unstable asymmetric mode in
the averaged equations is expected to correspond to an unstable periodic
solution of Eqs. (\ref{sys1}). In the bottom left panel of Fig.\ \ref{fig3},
we display the dynamics of the mode shown in the top right panel of the same
figure, represented by $u_{25}$ and $v_{25}$. It is seen that the
oscillating solution is indeed unstable, as predicted by the averaged system.

As another test of the accuracy of the approximation based on the averaging
method, in the bottom right panel of Fig.\ \ref{fig3} we display the
dynamics of a stable asymmetric mode with $f_{1}=6$ ($\sigma =4.15\times
10^{-5}$), as produced by direct simulations of the original equations (\ref%
{sys1}). Again, we observe a reasonable agreement between the two systems.

\section{Conclusion}

In this work, we have studied discrete solitons in the system of two
linearly coupled DNLSEs (discrete nonlinear Schr\"{o}dinger equations),
which describes several physical situations in BEC. For the case of the
coupling constant rapidly varying in time, we have derive the averaged
system of coupled generalized DNLSEs. In addition to the linear coupling,
the latter system contains nonlinear interaction terms of the
four-wave-mixing type. We have constructed families of symmetric single-site
localized states (fundamental discrete solitons), as well as in-phase and
twisted two-site modes, and identified their stability regions. In
particular, the two-site in-phase states are completely unstable, while
their twisted counterparts have a stability area. The symmetry-breaking
bifurcation of the fundamental discrete soliton, and the resulting
asymmetric fundamental localized mode have been found too, and the stability
of the asymmetric mode was explored. By means of direct simulations, it was
verified that, in all the cases, the averaged equations provide for a very
accurate approximation, in comparison with the underlying system featuring
the rapidly varying linear coupling.

\section*{Acknowledgments}

F.Kh.A. acknowledges a partial support through the Marie Curie IIF
grant. B.A.M. appreciates hospitality of the University of Hong Kong
and of the Hong Kong Polytechnic University.

\end{subequations}

\end{document}